\documentclass[10pt,a4paper]{article}
\usepackage{epsfig,amsmath}
\usepackage{color}
\numberwithin{equation}{section}
\numberwithin{equation}{subsection}
\usepackage{hyperref}
\def\A{\bar A}

\begin{document}
\title{Critical mass of neutron stars : a new view}
\author{Parthasarathi Majumdar\footnote{parthasarathi.majumdar@saha.ac.in}  \\A P C Division \\Saha
Institute of Nuclear Physics \\Kolkata 700064, India.} 
\maketitle
\begin{abstract}

The issue of the critical mass of neutron stars, with respect to gravitational
collapse to black holes, is reexamined from the perspective of thermal
stability of quantum horizons. Postulating the existence of a tiny, {\it
  embryonic}, isolated horizon, hidden deep inside a gravitationally
contracting neutron star, the critical
mass is seen to emerge from the extrapolation of the criterion of thermal
stability of quantum isolated horizons derived earlier by us, to such a
`hidden' horizon, as a
condition of its stability and growth (through formation of {\it trapping}
or {\it dynamical} horizons), eventually leading to an equilibrium
isolated horizon engulfing the entire star. The perspective is based on aspects of Loop Quantum Gravity,
and in contrast to extant analyses in the neutron star literature, uses neither classical spacetime metrics
nor details of strong neucleonic interactions at supranuclear densities,
thus delineating the essential role of quantum gravitation in black hole
formation. 

\end{abstract}

\section{Introduction}

The observed absence of neutron stars with masses in large excess of a few solar masses
strongly suggests the existence of a critical maximum mass for such stars. Heavier neutron
stars must become {\it unstable} with respect to gravitational collapse and
metamorphose into black holes. Since indirect evidence for black hole candidates with masses ranging
from a few to more than a billion solar masses have been widely reported,
the importance of understanding the critical mass of neutron stars can hardly
be overemphasized. Indeed, this has been an issue of some interest for over
three decades in neutron star research
\cite{mas,schu,yaz,vid,schu2,guo,prak,ruff}. Most approaches, with the
exception of \cite{ruff}  are
variations on the theme of derivation of the Chandrasekhar bound on
critical masses
of dying stars \cite{chandra} for which they may escape the white
dwarf stage. This standpoint entails hydrostatic equilibrium between
the gravitational pressure at the centre of the neutron star, on the one hand, and Fermi
degeneracy pressure due to the neutron star core modelled as a degenerate
relativistic neutron gas, on the other. This is invoked to determine the
highest critical mass that a neutron star can possess before it collapses to a
black hole due to gravitational predomination. More contemporary variations
focus on details of the equation of state of dense neutron (or quark) matter 
as appropriate to the inner regions of a neutron star. Since the density in
these regions exceeds nuclear densities ($\sim 10^{14}$ gm/cc), one is compelled to model the strong
interaction between nucleons and also thresholds of production of hyperons and
other resonances, through low energy effective field theory models of strong interaction dynamics at high
density. The equation of state to be used to
ascertain the Fermi degeneracy pressure is therefore rather sensitive to the
details of the low energy effective model employed, and therefore 
susceptible to the ambiguities inherent in its extraction from QCD, for instance. Perhaps
such ambiguities are unavoidable to some extent at the present state of development of
low energy strong interaction physics, if a sufficiently detailed knowledge of
the equation of state of dense nucleonic matter is desired. One could question
though to what extent such details are actually necessary for the estimate of a stability
upper bound on the mass of a neutron star against gravitational
collapse.  

Furthermore, observe that the low energy
effective strong interaction models employed to determine the equation
of state in the inner regions of a neutron star are
all formulated in a {\it flat Minkowski} spacetime background. In contrast, the gravitational
pressure  causing collapse is computed within {\it general relativity} using the
Tolman-Oppenheimer-Volkoff equation. When considering hydrostatic
equilibrium, one might worry that this obvious inconsistency in the
approach may adversely affect the final result. In other words, 
consistency demands that models to determine the equation of state be
formulated in accord with principles of {\it general relativistic} quantum field
theory. This challenging task has not really been taken up in earnest as far as I know.    
Under the circumstances, one may fear that even the best Minkowski space determination of the
equation of state may continue to suffer from large errors due to
neglect of spacetime curvature deep inside a neutron star. It is quite
remarkable however that such large errors are apparently not
there. This situation is reminiscent of John Mitchell's derivation of
the Schwarzschild radius of a gravitating sphere more than a century 
before Schwarzschild's own derivation, or more prominently, Niels
Bohr's derivation of the Bohr radius of the hydrogen atom, longer than
a decade prior to the
advent of quantum mechanics, or even Hans Bethe's estimate of the Lamb
Shift in hydrogen prior to the full Lorentz-covariant quantum
electrodynamics calculation
by French and Weiskopff and Feynman. In each of these examples,
though, the antiquated methodology is supplemented by some novel physical
insights which point to the correct theory developed in due course. It
is an interesting question as to which theory Chandrasekhar's approach
is a pointer to. 

In any event, a systematic approach to the problem of the maximum mass
of a stable neutron star should involve {\it general relativistic} quantum
field theory rather than special relativstic, and this is well-known to be a
challenging problem even in its formulation. One of the key
problems of quantum field theory in curved spacetime is that standard
results of special relativistic quantum field theory, like the spin
statistics theorem, which we take for granted in flat spacetime,
need to be established anew, that too on a
case-by-case basis, depending on the background geometry \cite{wald}.  

We have already mentioned above that ref. \cite{ruff} presents an exception to the
standpoint of most extant approaches. In this work, a maximum
mass of the neutron star has been derived based on the gravitational
hydrostatic pressure obtained from the Tolman-Oppenheimer-Volkoff equation,
the requirement that there is no local spontaneous collapse (Le Chatelier's Principle) and the
special relativistic requirement that the sound velocity is less than the speed of
light in vacuum. Thus, this approach is insensitive to the details of the
microscopic model of strong nucleonic interactions employed to determine the
equation of state of dense neutron matter. However, it does crucially use the
spherically symmetric ansatz of the classical spacetime metric in the
Tolman-Oppenheimer-Volkoff equation. 

In this paper we offer an alternative perspective, which is not only quite
insensitive to details of the strong interaction dynamics at high density, but
in fact also does not depend on the details of the classical spacetime geometry (metric)
inside the star. Thus, the paper may be considered to be an attempt to go
beyond ref. \cite{ruff}. We begin with the observation that the stability
upper bound for a neutron star can be expressed as an {\it instability
  lower} bound, i.e., the minimum mass of a neutron star for which it
will be unstable to gravitational collapse to a black hole
\begin{eqnarray}
\frac{M_{crit}}{M_P} > \xi ~ \left( \frac{\lambda_{Cn}}{l_P} 
\right)^2 = \xi ~ \frac{A_{Cn}}{A_P}  \label{instab}
\end{eqnarray}
where, $M_P~,~l_P~,~A_P~$ are respectively the Planck mass, length and
of the area, $\xi$ is a dimensionless constant of $O(1)$ and $\lambda_{An}$ is the Compton
wavelength of the neutron, with $A_{Cn} = \lambda^2_{Cn}$. Clearly, the right hand side of
(\ref{instab}) is {\it non-perturbative} in the Planck length $l_P$, as
distinct from perturbative quantum gravity effects which are
$O(l_P^n)~,~n \in {\bf Z}$. In fact, it is reminiscent of black hole
entropy \cite{kaul-sigma11, pm09} 
\begin{eqnarray}  
S_{bh} = \frac{A_{hor}}{4 A_P} + {\rm quantum~corrections} ~ \label{sbh}
\end{eqnarray}
which, likewise, arises from non-perturbative quantum spacetime
fluctuations. The question that arises then is : {\it Is the minimum
mass of a neutron star, for which it becomes unstable with respect to
gravitational collapse to a black hole, related in any way to the
entropy of the black hole ?} Unlikely as it may seem, the answer to
this question will turn out to be in the affirmative, of course
within a specific approach to quantum gravity, namely Loop Quantum
Gravity (LQG). The actual value of the critical mass that emerges from this relationship
lies within the correct order of magnitude, even though there is an ambiguity
of $O(1)$ which our approach cannot fix at the present state of
development. 

The paper is organized as follows: in the next section, we outline 
our proposal of formation of an isolated horizon from a
collapsing neutron star, starting with a hidden, `embryonic' horizon
(whose spatial foliation is an outer trapped 2-surface). This is
followed in section 3, by a digression into a sketch of our earlier derivation of a
thermal stability criterion of an equilibrium isolated horizon within
a canonical or a grand canonical ensemble of thermally active trapping
horizons, based on the notion of {\it Thermal Holography}. Even though most of
the material in this section is already published, its inclusion here is
primarily to make the paper self-contained. The 
stability criterion derived here, expressed as an inequality between the
mass of the isolated horizon and its {\it microcanonical} entropy (in
Planckian) units, is the key to our derivation of the critical mass of
a neutron star, as shown in section 4. We conclude in section 5 with a
discussion of our result and future outlook. 

\section{Horizon formation}

We begin with the postulate that the formation of a horizon (a
hypersurface in spacetime whose
spatial foliation is an outer trapped 2-surface) is not an abrupt
event, but rather an endpoint of a continuous succession of events,
beginning with the formation of a tiny, hidden (isolated) horizon deep inside
the collapsing neutron star and its subsequent growth. In this sense,
the eventual formation of an event or isolated horizon from a
collapsing neutron star is analogous to a first order phase transition
proceeding by bubble nucleation. The incipient 'embryonic' horizon forms by a
process presumably arising from quantum spacetime
fluctuations through a mechanism that is not yet known. Once the hidden baby
horizon has formed, its stability (and future growth) satisfies a certain thermal stability
criterion (to be elaborated upon in sections 3 and 4) which has been
derived \cite{cm04, pm07, mm11} on a far
sounder basis, from various aspects of loop quantum gravity
\cite{rov}. This stability criterion is expressed as a lower bound on
the mass of the hidden horizon in terms of its cross-sectional
area, in Planckian units. 

A pictorial representation is shown in Fig.s 1-3. 

\begin{figure}
\begin{center}
\hbox{
\hspace{1.5cm}
\includegraphics[height=6.0cm,width=6.0cm,angle=0]{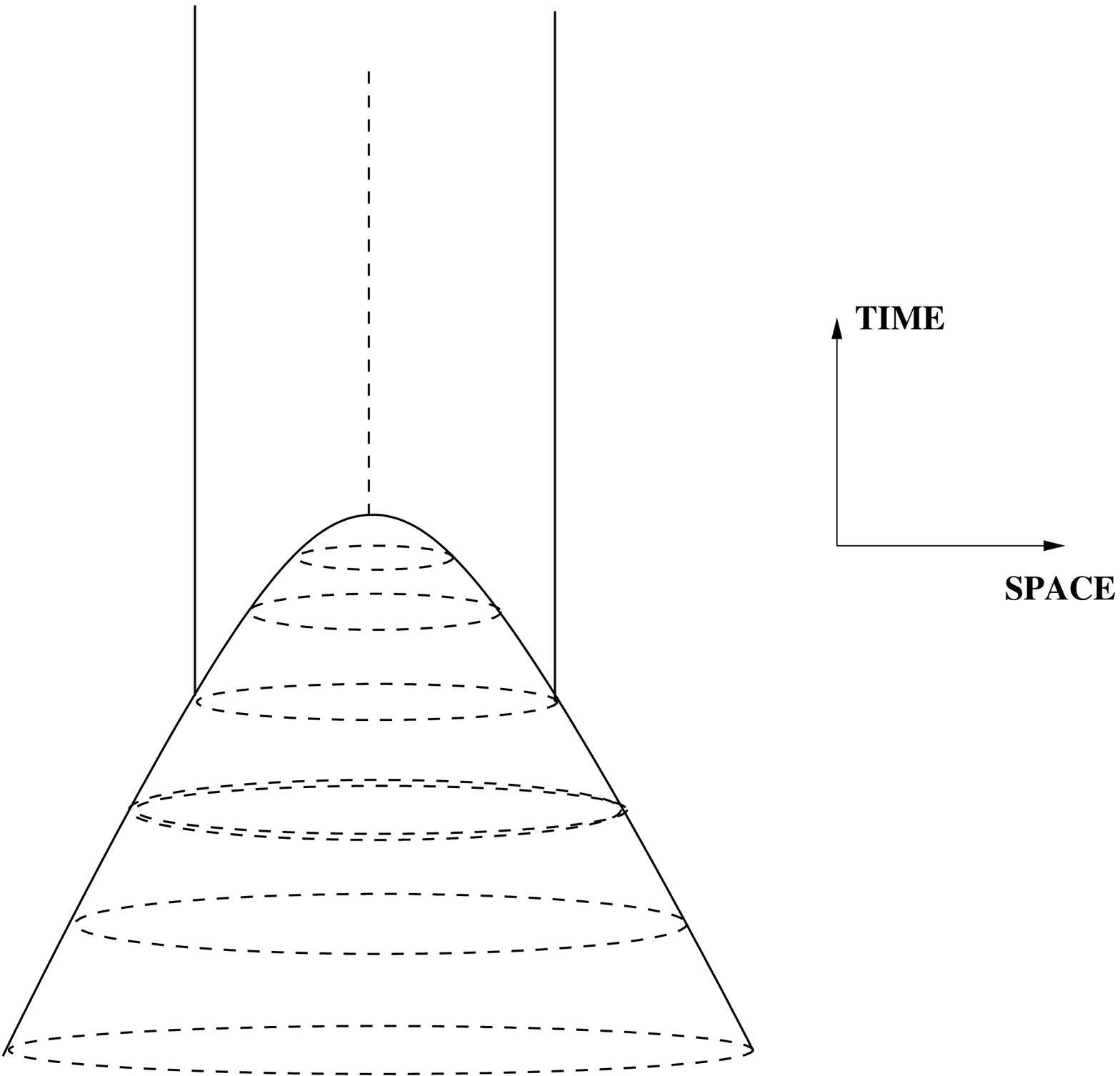}
\hspace{1.0cm}
\includegraphics[height=6.0cm,width=6.0cm,angle=0]{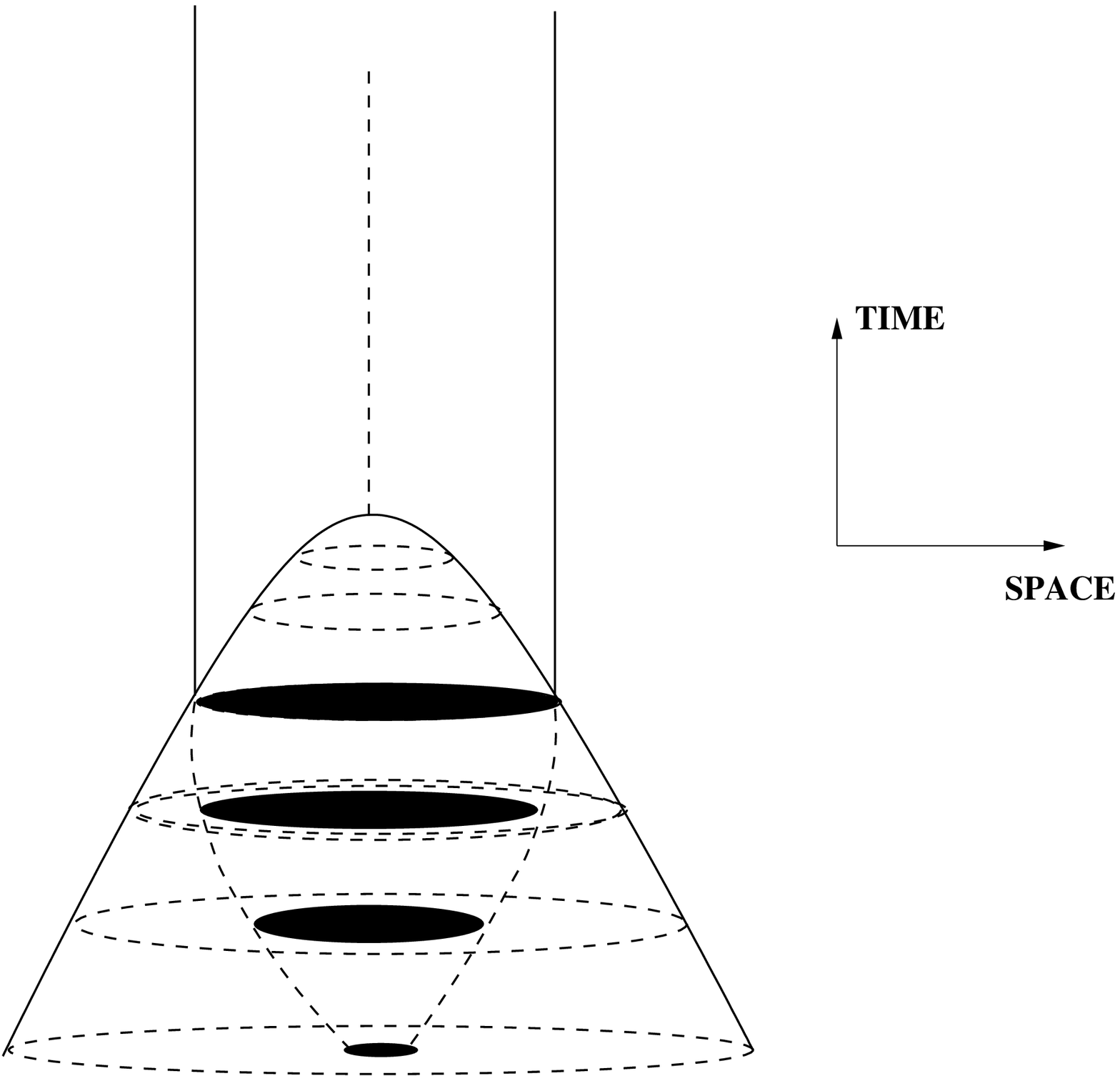}}
\end{center}
\vskip -.6cm
\caption{\footnotesize{{\it Left panel}: Gravitational collapse of a spherical star in
Eddington-Finkelstein coordinates.
{\it Right panel}: Growth of nascent hidden horizon inscribed within
collapsing star. Filled-in ellipses depict the hidden horizons at
various spatial foliations.}}
\label{fig1}
\end{figure}

In Fig.1, on the left panel a cartoon is shown of a spherically collapsing star 
in Eddington-Finkelstein coordinates; the broken ellipses 
depict the collapsing star and the vertical unbroken lines 
are the event horizons formed when the star collapses down 
to a certain size. The collapse continues beyond the 
formation of the horizon, all the way to a singularity, shown 
here by the broken vertical line.

%\begin{figure}
%\begin{center}
%\epsfxsize=6cm
%\epsfbox{/home/partha/talks/figs/critneum-2.eps}
%\begin{caption}{\small Growth of nascent hidden horizon inscribed within
%collapsing star}
%\end{caption}
%\end{center}
%\end{figure}

The right panel of Fig. 1 depicts a picture of gradual growth of the proposed nascent hidden 
horizon into the physical horizon which coincides with the standard 
conceptualization. The filled ellipses inscribed inside the collapsing star 
represented by the broken ellipses indicate this growth. Since, eventually 
the entire star is now eaten up by the growing horizon,  there is no 
more energy to be accreted from the star `outside', and the growth 
stops. Subsequent to this, the star continues to collapse, eventually 
reaching the black hole singularity.

Because of the teleological nature of event horizons of black holes
where a global timelike isometry is a requirement, we work with
isolated horizons which are characterized completely locally
\cite{asht-krish-lr} within a non-stationary spacetime. These are depicted in collapsing situations in
Fig. 2. The figure on the left shows two horizons forming in collapse,
the formation of the first being followed by an accretion leading to
the second. The figure on the right is the conformal depiction of the same
spacetime. Since the characterization of isolated horizons is local,
each of the horizons H and H' are equivalent isolated horizons with
different cross-sectional areas and masses. In the language of event
horizons, H would not qualify as an event horizon since it accretes
matter; H' might be taken to be an event horizon provided no further
accretion happens in future. But one needs to `know' that in advance
if one is to think of H' as an event horizon. One might expect to
avoid this teleological conundrum by using {\it apparent} horizons,
but then these latter are (a) described only on spatial foliations and
(b) change discontinuously under physical changes like accretion. The
second aspect stymies a description in terms of classical evolution in
phase space which is continuous, and makes apparent horizons less suited for
our intended formulation in terms of local variables. Isolated
horizons afford such local descriptions. Briefly, they are
characterized as follows :
\begin{itemize}
\item Isolated horizons are {\bf Nonstationary}, i.e., are null boundaries of
spacetimes which are not required to have global timelike isometries.
\item An isolated horizon is a Null (lightlike) inner boundary of sptm with
topology $R \otimes S^2$. 
\item It is {\it marginally} Outer Trapped : in terms of the null geodesic
generators $l$ and $n$, $\theta_{(l)} = 0~,~\theta_{(n)}< 0$.
\item An isolated horizon has cross-sectional area ${\cal A}(S^2) = const
\rightarrow$ meaning of isolation. 
\item {\it Zeroth law of Isolated Horizon Mechanics :} surface gravity
$\kappa_{IH} = const$ {\it on} isolated horizon, even though elsewhere it may
not be possible to define this quantity due to lack of stationarity.
\item It is possible to define {\it mass} on an isolated horizon : $M_{IH} = M_{IH}(A,Q)$
where, $M_{IH} \equiv M_{ADM} - {\cal E}_{rad}^{\infty} $, where $M_{ADM}$ is
the ADM mass,  such that 
$ \delta M_{IH} = \kappa \delta {\cal A}_{hor} + \Phi \delta Q_{hor}$ {\it
  (First law of Isolated Horizon Mechanics)}
\item IH is {\it microcanonical ensemble} with fixed $ {\cal A}_{hor}, Q_{hor}  $
\item Hawking radiation requires IH $\rightarrow$ Trapping or Dynamical Hor
\end{itemize}

\begin{figure}
\hbox{
\hspace{4.0cm}
\includegraphics[height=6.0cm,width=9.0cm,angle=0]{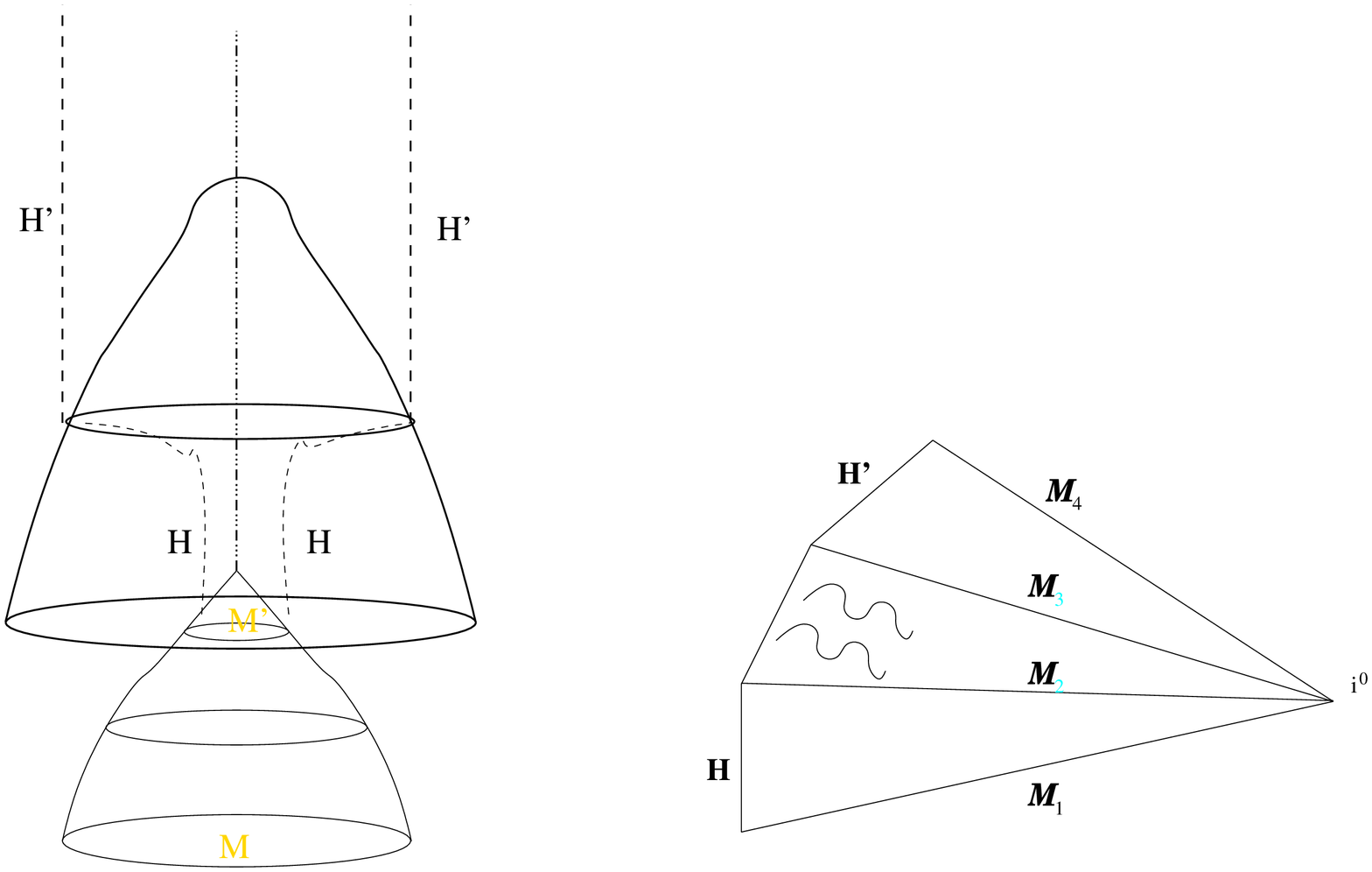}}
\caption{\footnotesize{Isolated Horizons}}
\label{fig2}
\end{figure}
  
The nascent, stable, hidden isolated horizon stage is
followed by a stage of a hidden trapping/dynamical horizon \cite{haywd97,
asht-krish-lr} which accretes energy (matter) and grows, and then
settles down to a bigger hidden isolated horizon, still substantially inside
the neutron star inner region. This isolated horizon changes to a
trapping horizon and grows via accretion, subject to its fulfilling our stability
criterion. This process of alternation between the hidden isolated and expanding
trapping horizons continues
until the entire star is engulfed by the isolated horizon which is
then identified with {\it physically observed} horizon. The neutron star of course
continues to shrink inside the horizon until it reaches the
singularity. This proposed scenario is depicted in Fig. 4 in the
Carter-Penrose frame. 

\begin{figure}
\begin{center}
\epsfxsize=6cm
\epsfbox{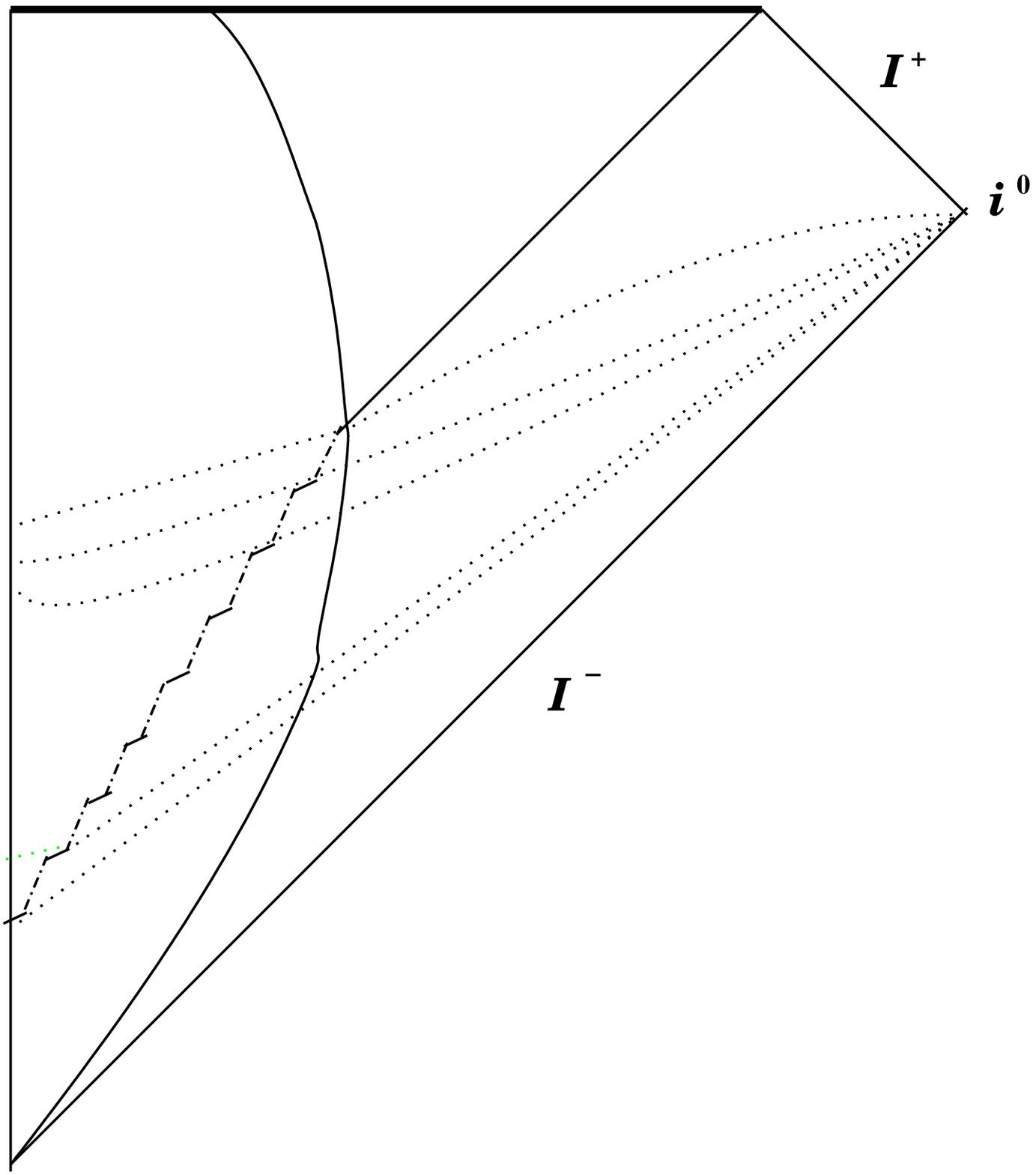}
\begin{caption}{\footnotesize{Proposed scenario for growth of hidden horizon in
spherical collapse. Figure shows alternate isolated and trapping
horizons culminating in emergence of physical isolated horizon. The
part of the collapsing star to the left of the growing hidden horizon,
at any spatial foliation (indicated by dotted lines)
is to be taken to be {\it trapped}.}} 
\end{caption}
\end{center}
\end{figure}

\section{Digression}  

\subsection{Thermal Holography}

In any version of quantum general relativity (QGR), the Hamiltonian
constraint of the classical theory would be exhibited as the
annihilation of bulk quantum states $|\psi_v\rangle$ (consisting of both the
quantum bulk spacetime and matter states) by the bulk Hamiltonian
operator, assuming that this latter exists,
\begin{eqnarray}
{\hat H}_v ~|\psi_v\rangle ~=~ 0 ~. \label{hcon}
\end{eqnarray}
One can interpret (\ref{hcon}) as the {\it quantum} Einstein
equation. While classically, properties of the bulk spacetime determine
completely properties of the boundary, in QGR this need not be the
case. Thus, one may assume that the full Hilbert space ${\cal H} \sim
{\cal H}_v \otimes {\cal H}_b$ where ${\cal H}_b$ is the Hilbert space
of quantum states describing boundary fluctuations. An arbitrary state
in this Hilbert space may thus be expressed as
\begin{eqnarray}
|\Psi \rangle ~ = ~ \sum_{v,b} c_{vb} \underbrace{
 |\psi_v\rangle}_{{blk}} \underbrace{|\chi_b \rangle}_{{bdy}} ~, \label{anys}
\end{eqnarray}
where we have assumed minimal entanglement between the bulk and the
boundary, i.e., the coefficient matrix $c_{vb}$ is not necessarily
diagonal. 

Consider now a canonical ensemble of such spacetimes in contact with a
heat bath with an inverse temperature $\beta$. The canonical partition
function 
\begin{eqnarray}
Z &=& \sum_{{}_{b}} \left( \sum_{{}_v} |c_{{}_{vb}}|^2 ||
  ~|\psi_{{}_v}\rangle~||^2
\right) \langle \chi_{{}_b}|\exp - \beta {\hat H}_{{}_{bdy}} |\chi_{{}_b}
\rangle \nonumber \\
& \equiv & Z_{{}_{bdy}} ~,~\label{thol} 
\end{eqnarray}
using the quantum Einstein equation for the bulk states. In other
words, {\it the bulk states decouple} because of the quantum Einstein
equation, and {\it thermodynamics (of any quantum spacetime) is completely
described by the boundary states} \cite{pm09, pm07}! We term this phenomenon {\it
thermal holography}; admittedly this is weaker result than the
assertions in the {\it Holographic Hypothesis} \cite{thft, suss, bou},
but a result nevertheless, albeit heuristic, rather than a conjecture. 

\subsection{Thermal Stability}

Consider now the canonical ensemble above to consist of (inner) boundaries
which are trapping or dynamical horizons which are thermally
active. The result of the previous subsection, that the partition function is described in terms
of the boundary states, can be interpreted, within the approach to
quantum gravity known as Loop Quantum Gravity (LQG) \cite{rov}, to imply
that the quantum numbers labelling the states must correspond to some
observable within LQG associated with such boundaries. One of the most
important such observable is the area operator. Even though this
operator is generically not a Dirac observable \cite{rov} in the strictest sense, the
area operator corresponding to {\it physical} surfaces like
(spatial foliations of) horizons are indeed good observables. In LQG,
the spectrum of the area operator is discrete and bounded, 
\begin{eqnarray}
a_N &=& 8\pi \gamma A_P \sum_{p=1}^N \sqrt{j_p(j_p+1)} \label{aspec} \\
|a_N| &\leq& A_{hor} \pm O(A_P) ~, \label{bded}
\end{eqnarray}
where, the sum is over punctures on the horizon made by edges of spin
network states describing quantum spacetime. Now, for macroscopic
horizons whose areas $A_{hor} >> A_P$, one can consider for
simplicity, states where all spins at punctures are 1/2. In this
situation, $a_N \sim N~A_P~,~N >>1$. 

Continuing with the canonical ensemble considered in the previous
subsection, the boundary partition function corresponding to a
canonical enselble of thermally active trapping horizons can be reexpressed in
terms of functions of the area eigenvalues $a_N$
\cite{dmb01,cm04,pm07,mm11} 
\begin{eqnarray}
Z_b(\beta) = \sum_n g(M(a_n)) \exp - \beta M(a_n) ~, \label{zbdy}
\end{eqnarray}
where, we have introduced an unspecified function $M(a_N)$ of the
horizon area eigenvalues as the {\it mass} of the horizon. We now
assume that the ensemble has an equilibrium configuration which is an
{\it isolated} horizon as defined in section 2. We also assume that the partition
function (\ref{zbdy}) can be evaluated in saddle point approximation including
Gaussian fluctuations around the saddle point, taken to be an isolated
horizon of cross-sectional area $A_{IH}$. Changing the sum to an
integral for large horizon areas ($N >> 1$), we obtain
\cite{dmb01, cm04, pm07}
\begin{eqnarray}
Z_b(\beta) \simeq & \exp \left[S(A_{IH}) - \beta M(A_{IH}) \right]
\cdot \Delta^{-1/2}(A_{IH}) ~, \label{sadpt}
\end{eqnarray}
where, $S(A_{IH}) \equiv \log g(A_{IH})$ is the {\it microcanonical}
entropy of the equilibrium isolated horizon, and 
\begin{eqnarray}
\Delta(A_{IH}) \equiv M^{-1}_A ~ \left[ M_{AA} S_A - S_{AA} M_A
\right] ~, \label{Del}
\end{eqnarray}
where the subscripts $A$ imply derivatives of these functions with
respect to their argument and we have dropped the subscript $IH$ from
the $A_{IH}$ for convenience.

Evaluating the {\it canonical} entropy using standard thermodynamic
formulae, we get
\begin{eqnarray}
S_{canon}(A) = S(A) + \frac12 \log \Delta(A)
\end{eqnarray}
where, as already mentioned, $S(A)$ is the microcanonical entropy of
the equilibrium horizon taken as the saddle point, we see that, for
$S_{canon}$ to have the standard interpretation of entropy, we must
have $\Delta(A) > 0$. Integrating the corresponding differential
inequality, using (\ref{Del}), and reinserting the fundamental constants set to unity
earlier, we get the condition
\begin{eqnarray}
\frac{ M(A_{IH})}{M_P} > \frac{S(A_{IH})}{k_B} ~ .\label{stab}  
\end{eqnarray}
We take this condition to be the {\it necessary and sufficient condition}
for the equilibrium isolated horizon $IH$ to be thermally stable
\cite{cm05, pm07}. If this inequality is violated, the isolated
horizon is not a stable equilibrium configuration and the resulting
instability leads to an incessantly radiating (or thermally accreting) trapping
horizon, which may eventually evaporate and disappear (in the case of radiation). A
similar thermal stability criterion has been derived recently for {\it charged}
trapping horizons \cite{mm11}. 

The stability criterion (\ref{stab}) depicts the domination or otherwise of
processes that are `energy (mass)-driven' rather than `entropy-driven'. In
other words, in the context of the collapsing neutron star, the domination
of entropy-driven processes may be considered tantamount to the domination of
Fermi pressure due to Pauli exclusion in the neutron core, which may well
stave-off gravitational collapse to a black hole. The concept of a hydrostatic
equilibrium between `forces' due to gravity and that due to Fermi pressure of
degenrerate neutrons is replaced here by an `equilibrium' betweem the horizon
mass and its microcanonical entropy - both statistical concepts rooted in
quantum aspects of spacetime. 
 
\subsection{Microcanonical Entropy}

One observes that the non-triviality of the stability criterion (\ref{stab}) is
crucially dependent on the microcanonical entropy being a more general
function of the horizon area than merely linear (i.e., merely
Bekenstein-Hawking). Indeed, the Bekenstein-Hawking area law for black
hole entropy was historically derived using {\it semiclassical}
arguments where a classical spacetime metric is used. It is thus
subject to modifications within a genuine proposal for {\it quantum}
spacetime geometry like LQG. Indeed, longer than a decade ago, the
microcanonical entropy of spherical isolated horizons was first
investigated \cite{ack98, km98, km00, abck00}, on the basis of
counting states of an $SU(2)$ Chern Simons theory describing the dynamics on
the horizon, as derived using LQG. Using the connection \cite{wit86}
between the Chern Simons theory Hilbert space on a
3-manifold and the conformal current blocks of the Wess-Zumino-Witten
model on the boundary of the 3-manifold, the microcanonical entropy
was shown, for macroscopic ($A_{IH} >> A_P$) horizons, to be given by
an infinite series, asymptotic in the horizon area \cite{km00, dkm01},
beginning with the BH area law, but with calculable and finite
corrections :
\begin{eqnarray}
S(A_{IH}) = S_{BH}(A_{IH}) - \frac32~\log S_{BH}(A_{IH}) + const +
O((S_{BH}^{-1})) ~, \label{kms}
\end{eqnarray}
where, $S_{BH}(A_{IH}) \equiv A_{IH}/4 A_P~,~k_B=1$ is the BH area law for
entropy. Because of the logarithmic corrections (and beyond) in (\ref{kms}), $S_{AA}$ is
nontrivial, leading to the form of the stability criterion given in
(\ref{stab}). This underlines the essentially quantum gravitational nature
of thermal stability of black holes, since, instead of using the classical
metric and properties derived from classical geometry, LQG aspects have been
germane to the derivation of (\ref{stab}). 

\section{Critical Mass}

Assume that the fractional energy (mass) loss of the collapsing neutron star
is small, such that the original mass of the star is more or less completely
engulfed by the horizon, 
\begin{eqnarray}
\frac{M_{crit}}{M_P} > \xi~ \frac{M_{IH}}{M_P} ~. \label{mmc}
\end{eqnarray}
Recall our picture on the right panel
of Fig. 1 where, a nascent `embryonic' quasi-isolated horizon which is stable
according to our criterion (\ref{stab}), changes to a trapping horizon
$TH_0$ which accretes neutron-rich material from its environment and grows. Thus,
one has a series of alternate (quasi-) isolated and trapping horizons in the
sequence, pictorially depicted in Fig. 3.  
\begin{eqnarray}
IH_0 \rightarrow TH_1 \rightarrow IH_1 \rightarrow TH_2 \rightarrow \cdots
TH_N \rightarrow IH ~(physical) ~. \label{prop}
\end{eqnarray}
This sequence requires the stability conditions for the quasi-isolated
horizons $IH_0, \dots, IH_N$
\begin{eqnarray}
\frac{M_{IH_i}}{M_P} > \frac{S_{IH_i}}{k_B}~,~i=0,1,2,...,N~\label{stabi}
\end{eqnarray}
implying the sequence of inequalities 
\begin{eqnarray}
A_{IH_0} < A_{IH_1} < A_{IH_2} < ... < A_{IH} ~(physical) ~. \label{ineq}
\end{eqnarray}
Observe that if any of the inequalities (\ref{stabi}) fails to hold, this
implies that the hidden quasi-isolated horizon is unstable, and will no longer change
to a trapping horizon which will accrete and grow. Rather, the nucleation
process stops there and the hidden horizon simply dissolves into the neutron
star core. 

If indeed, as assumed, $IH_0$ is the incipient quasi-isolated horizon, then $A_{IH_1}$
is the cross-sectional area of the isolated horizon which the trapping horizon
$TH_1$ has settled into. In this case, since the foliation of $TH_1$ is an
outer trapping surface, $A_{IH_1} > A_{Cn}$. The sequence of inequalities
(\ref{ineq}) then implies 
\begin{eqnarray}
\frac{A_{IH}}{A_P} > \frac{A_{Cn}}{A_P} ~. \label{crith}
\end{eqnarray}
Recalling the stability criterion applied now to the {\it actual} isolated
horizon
\begin{eqnarray}
\frac{ M(A_{IH})}{M_P} > \frac{S(A_{IH})}{k_B} = \xi ~\frac{A_{IH}}{A_P} ~ ,\label{stab}  
\end{eqnarray}
and using eq.s (\ref{crith}) and (\ref{mmc}), we obtain
\begin{eqnarray}
\frac{M_{crit}}{M_P} > \xi~\frac{A_{Cn}}{A_P} ~\label{critm}
\end{eqnarray}
from which eq. (\ref{instab}) follows immediately. Thus, upto the $O(1)$
constant $\xi$, the minimum mass of a collapsing neutron star for which it
will collapse to a black hole, obtained from the above analysis, has the
correct order of magnitude in terms of fundamental constants (or the solar
mass).     

\section{Discussion}

The formulation of the problem of critical mass of a neutron star beyond
which it becomes unstable with respect to gravitational collapse to a black
hole, has been made here in a {\it dual} fashion, namely, by investigating the {\it
  minimum} mass for which such a gravitational collapse occurs, with the
concomitant formation of an event (isolated) horizon. Thus, the focus has been
on the description of horizon formation rather than the instabilities of a
neutron star brought about by strong interaction dynamics of a dense neutron gas in
its interior, in a background of strong gravity. This shift of focus has enabled
the facile use of results pertaining to horizon dynamics and thermodynamics
which result from various aspects of quantum gravity, especially LQG. The
complications associated with the traditional approach regarding possible
inconsistencies in using classical {\it general} relativity to determine the
hydrostatic pressure in the neutron star interior, while using an equation of
state determined from {\it special} relativistic quantum field theory, have been
thus completely obviated. Further, this formulation underlines the
{\it quantum} gravitational underpinnings of the process of horizon
formation, in case gravitational collapse does occur. This is a novel
feature of our assay in comparison with that of ref. \cite{ruff} where
the incipient predominance of gravitational effects over the strong
dynamics of densely packed neutrons is already in evidence. However,
it is not possible to obtain a thermal stability criterion like we
have derived, on the basis of classical general relativity.   

Indeed, horizon thermodynamics is inherently quantum to the extent that
quantum gravity affords the only sensible ab initio description of notions like black hole
entropy \cite{pm09, kaul-sigma11}. Even the entropy of galactic centre
black holes weighing billions of solar masses has its origin in quantum
gravity, i.e., microstates describing physics at $10^{-33}~cm$. Black holes
must therefore be an extreme example of macroscopic quantum phenomena. While
the entropy of isolated black holes is now reasonably well-understood on the
basis of quantum gravity (e.g., LQG), a myriad aspects of quantum black holes are yet
to be understood in detail, chief among which is the proper understanding of
black hole singularities. This aspect is of crucial importance if one is to
resolve the so-called Information Loss Paradox. On our part, the thermal
stability of black holes has been a subject of earlier investigations, and our result from that
study appears
to yield the critical mass more or less straightforwardly. Nevertheless, there
are features of our analysis which are heuristic in character, and these
features need thorough reexamination before any claim of rigour can be made. 

Neutron stars are currently detected as
sources of moderately hard X-rays, with tell-tale accretion discs in case they
are a partner in an X-ray binary system. It is not easy at all to decipher from
X-ray data (or even from radio frequency data emitted by pulsars) information regarding
the degrees of freedom and dynamics in the interior of neutron stars. The
somewhat unexpected link
seen here between the critical mass and the entropy of the nascent hidden
horizon emerges from a proposed scenario of horizon formation by a process of `bubble'
nucleation as in a first order phase transition. At this stage of
development, it is far from clear what would constitute an
observational evidence for this phase transition. There are models of
neutron star structure where certain layers close to the outer
boundary may have superfluid or superconducting properties, and
there are expectations that a careful study of the X-ray spectrum may
yield observational evidence for this. In this vein, it is possible
that the hidden horizon growth discussed in the text of the paper may
also radically influence the X-ray spectrum towards the later stages
of the collapse.

Finally, we note in passing that in the classical description of gravitational collapse of spherical
pressureless dust by J. R. Oppenheimer and H. Snyder, or that of a spherical
distribution of electromagnetic radiation by P. C. Vaidya, the growth of a trapped surface deep inside the
collapsing matter or radiation may actually be studied as toy models within classical
general relativity. We hope to report on this in the near future \cite{chm}.   

\noindent {\bf Acknowledgement:} I thank D. Bandyopadhyay for
illuminating discussions and for guiding me to the literature on
neutron stars.


\begin{thebibliography}{99}
\bibitem{mas} E. Massot, J. Margueron and G. Chanfray, Europhys. Lett. {\bf
97}, 39002, 2012; e-Print: arXiv:1201.2772 [nucl-th].
\bibitem{schu} H.-J. Schulze, T. Rijken, Phys.Rev. {\bf C84}, 035801, 2011.
\bibitem{yaz} T. Yazdizadeh and G. H. Bordbar, Res. Astron. Astrophys.{\bf
11} 471, 2011; e-Print: arXiv:1012.2939 [astro-ph.SR]
\bibitem{vid} I. Vidana, D. Logoteta, C. Providencia, A. Polls, I. Bombaci,
Europhys. Lett. {\bf 94} 11002, 2011; e-Print: arXiv:1006.5660 [nucl-th].
\bibitem{schu2} H.-J. Schulze, A. Polls, A. Ramos and I. Vidana,
Phys. Rev. {\bf C73}, 058801, 2006.
\bibitem{guo} G-Q Li, C.H. Lee, G.E. Brown, Phys. Rev. Lett., {\bf 79}, 5214, 1997.
e-Print: nucl-th/9711002
\bibitem{prak} M. Prakash, T. L. Ainsworth and J. M. Lattimer,
Phys. Rev. Lett. {\bf 61}, 2518, 1988.
\bibitem{ruff} C. E. Rhoades and R. Ruffini, Phys. Rev. Lett., {\bf 32}, 324, 1974.
\bibitem{chandra} S. Chandrasekhar, Astrophy. Jou. {\bf 74}, 81, 1931. See also,
S. Chandrasekhar, {\it Nobel Lecture (Physics) 1983} in {\it
Nobel Lectures, Physics 1981-1990}, Chief Ed. T. Fr\"angsmyr, Editor
G. Ekspang, World Scientific Publishing Co., Singapore, 1993.
\bibitem{wald} R. M. Wald, {\it Quantum Field Theory in Curved Spacetime and
Black Hole Thermodynamics}, Chicago, 1995. 
\bibitem{kaul-sigma11} R. K. Kaul, SIGMA {\bf 8}, 005, 2012. e-Print: 1201.6102 [gr-qc].
\bibitem{pm09} P. Majumdar, Int. Jou. Mod. Phys. {\bf A24}, 3414,
  2009. e-Print: arXiv:0903.5080 [gr-qc].  
\bibitem{cm04} A. Chatterjee and P. Majumdar, Phys. Rev. Lett. {\bf 92}, 
141302, 2004. e-Print: gr-qc/0309926.
\bibitem{pm07} P. Majumdar, Class. Quant. Grav. {\bf 24}, 1727, 2007. e-Print: gr-qc/0701014.
\bibitem{mm11} A. Majhi and P. Majumdar, {\it Charged Quantum Black Holes: Thermal Stability Criterion},
e-Print: arXiv: 1108.4670 [gr-qc].
\bibitem{rov} C. Rovelli, {\it Quantum Gravity}, Cambridge, 2004.
\bibitem{asht-krish-lr} A. Ashtekar and B. Krishnan, Liv. Rev. Rel. {\bf 7},
10, 2004. 
\bibitem{haywd97} S. A. Hayward, Phys. Rev. {\bf D49}, 6467, 1994. 
\bibitem{thft} G. 't Hooft, {\it Dimensional Reduction and Quantum Gravity},
  in {\it Salamfestschrift}, edited by A. Alo and S. Randjbar-Daemi, ICTP,
  Trieste, 1993. 
\bibitem{suss} L. Susskind, Jou. Math. Phys. {\bf 36}, 6377, 1995. e-Print: hep-th/9410074.
\bibitem{bou} R. Bousso, Rev. Mod. Phys. {\bf 74}, 825, 2002. e-Print:
  hep-th/0203101. 
\bibitem{dmb01} S. Das, P. Majumdar, R. K. Bhaduri, Class. Quant. Grav. {\bf
19}, 2355, 2002. e-Print: hep-th/0111001,
\bibitem{cm05} A. Chatterjee and P. Majumdar, Phys. Rev. {\bf D72} 044005, 2005.
e-Print: gr-qc/0504064.
\bibitem{ack98} A. Ashtekar, J. Baez, A. Corichi and K. Krasnov,
  Phys. Rev. Lett. {\bf 80}, 904, 1998. e-Print: gr-qc/9710007.
\bibitem{km98} R. K. Kaul and P. Majumdar, Phys. Lett. {\bf B439}, 267,
  1998. e-Print: gr-qc/9801080.
\bibitem{km00} R. K. Kaul and P. Majumdar, Phys. Rev. Lett. {\bf 84}, 5255, 2000.
e-Print: gr-qc/0002040.
\bibitem{abck00} A. Ashtekar, J. Baez and K. Krasnov,
  Adv. Theor. Math. Phys. {\bf 4}, 1, 2000. e-Print: gr-qc/0005126. 
\bibitem{wit86} E. Witten, Commu. Math. Phys. {\bf 121}, 251, 1989.
\bibitem{dkm01} S. Das, R. K. Kaul, P. Majumdar, Phys. Rev. {\bf D63}, 044019,
2001. e-Print: hep-th/0006211.
\bibitem{chm} C. Chakraborty and P. Majumdar, in preparation.
\end{thebibliography}
\end{document}